\documentclass[twocolumn,showpacs,preprint2,amsmath,amssymb,superscriptaddress,floatfix]{revtex4}

\usepackage{epsfig}
\usepackage{bm}

\begin{document}

\title{Fission Cycling in Supernova Nucleosynthesis: Active-Sterile Neutrino Oscillations}

\author{J. Beun}
\affiliation{Department of Physics, North Carolina State University, Raleigh, NC 27595-8202}
\email{jbbeun@unity.ncsu.edu} 
\author{G. C. McLaughlin}
\affiliation{Department of Physics, North Carolina State University, Raleigh, NC 27595-8202}
\author{R. Surman}
\affiliation{Department of Physics, Union College, Schenectady, NY 12308}
\author{W. R. Hix}
\affiliation{Physics Division, Oak Ridge National Laboratory, Oak Ridge, TN 37831-6374}

\date{\today}

\begin{abstract}  

We investigate nucleosynthesis in the supernovae post-core bounce neutrino-driven wind environment
in the presence of active-sterile neutrino transformation.
We consider active-sterile neutrino oscillations for a range of mixing parameters: vacuum mass-squared differences of 
$0.1 \, \rm{eV}^2 \leq$ $\delta m^2 \leq 100 \, \rm{eV}^2$, 
and vacuum mixing angles of  $\sin^2 2  \theta_v \geq 10^{-4}$.  
We find a consistent $r$-process pattern for a large range of mixing parameters that is in rough agreement with the halo star CS 22892-052 abundances
and the pattern shape is determined by fission cycling.
We find that the allowed region for the formation of the $r$-process peaks overlaps the LSND and NSBL (3+1) allowed region.  

\end{abstract}


\maketitle

\section{Introduction}

Neutrinos dominate both the dynamics and the nucleosynthesis in the region near the protoneutron star during a
core collapse supernova. The prodigious neutrino flux generated here is thought to deposit enough energy to 
liberate material from the 
gravitational well of the protoneutron star.
Neutrino capture produces a neutron-rich environment outside the protoneutron star
at late times post-core bounce.
Understanding the role of neutrinos in this
environment directly leads to our understanding of heavy element nucleosynthesis in the core collapse supernova environment.

One of the synthesis processes that may take place in this environment is the $r$process, or rapid neutron capture process, 
which generates about half of the elements with atomic weight $A \geq 100$ \cite{Burbidge et al.(1957)}.  
At the heart of the $r$process is the rapid capture of neutrons, which occurs much faster than the
competing beta decay, forming very neutron rich nuclides.  As the supply of free neutrons is exhausted, these nuclides 
decay back to beta stability, forming a characteristic $r$-process abundance pattern that can be observed today.

While this basic $r$-process mechanism is understood, finding the astrophysical location of the $r$process
proves more elusive.  Several observational factors point towards the neutron-rich material produced near the protoneutron star
of a core-collapse supernovae as a likely candidate, \emph{e.g} \cite{Sneden Cowan(2003),Qian et al.(1998)} .  
Sneden and Cowan 2003 \cite{Sneden Cowan(2003)} concluded the $r$-process abundance patterns in extremely metal poor giant
stars match the second and third peaks of the solar system $r$-process pattern, indicating these elements were formed early in the evolution
of the universe given these considerations \cite{Argast:2003he}.  The neutrino-driven wind of the core collaspe
supernova is a promising candidate for the $r$process, for a review see Wanajo et al. 2001 
\cite{Wanajo et al.(2001)}.

The evolution for a mass element which will eventually undergo an $r$process in the neutrino-driven wind of the protoneutron star proceeds
 through several stages \cite{Takahashi et al.(1994),Woosley:ux,Meyer et al.(1992),Qian et al.(1993),Thompson et al.(2001)}.
Material first emerges from the surface of the protoneutron star as free nucleons, carried off by the neutrino-driven wind.
The strong influence of neutrinos in this regime produces neutron rich material resulting from the equilibrium 
effects of the $\nu_e$ and $\overline\nu_e$ fluxes, as the $\overline\nu_e$'s are thought to have a longer mean free path at the surface of
 the protoneutron star and hence are more energetic.
As the mass element moves farther from the star, it reaches lower temperatures ($T < 750$ keV), the nucleons coalesce into $\alpha$ particles and a large
 abundance of free neutrons.  The $\alpha$ particles then combine into seed nuclei for the $r$process with 
$50 \lesssim A \lesssim 100$.  Neutron capture begins at even lower 
temperatures ($T < 300$ keV) allowing the formation of $r$-process elements.

It has long been suggested that fission could influence the $r$process \cite{Seeger et al.(1965)}, primarily 
for cosmochronometers \emph{e.g} \cite{Cowan et al.(1987),Cowan 1999,Chechtkin et al. 1988}
In sufficiently neutron-rich conditions, heavy elements which are unstable to fission can be produced,
terminating the $r$-process path.
Additionly, fission influences the $r$process through the subsequent cycling of these fission unstable heavy elements
by neutron capture on the fission products.  
Beta-delayed, neutron induced, and spontaneous fission are all thought to play important roles in determining the outcome
of the final $r$-process abundances \cite{Panov Thielemann(2003)}.

Early $r$-process models in the neutrino-driven wind environment, such as Woosley, et al, 1994 \cite{Woosley:ux}, accurately
 reproduced the $r$-process abundances without significant fission cycling;
However, later models produced lower entropy per baryon, producing nearly
but not quite all the requisite conditions.  Further, proper consideration
of the near inertness of alpha particles to neutrino interactions resulted
in decreased neutronization, presenting a significant impediment to 
$r$-process production in this environment
\cite{Meyer(1995),Fuller Meyer(1995),Hoffman et al.(1997),Takahashi et al.(1994),McLaughlin:1997qi,Meyer:1998sn}.

There are three possible solutions for circumventing these problems.  One solution is 
modifying the hydrodynamics by using a 
very fast dynamical timescale \cite{Hoffman et al.(1997),Cardall Fuller(1997)}
even in a proton-rich environment \cite{Meyer(2002)}, or increasing temperature with high entropy \cite{Woosley:ux}; 
however, it is not known how neutrino heating 
could generate such conditions.
Another possibility is choosing a different location for the $r$process to occur, such as a neutron-star merger
\cite{Meyer(1989),Rosswog et al.(1999)}.
However this solution is not currently favored by analysis of the observational data \cite{Argast:2003he}.
Other astrophysical sites such as gamma ray burst accretion disks are currently being considered
\cite{Surman McLaughlin(2004),McLaughlin:2004be}.

The solution examined here is the introduction of active-sterile neutrino oscillations through the
$\nu_e \leftrightarrows \nu_s$ and $\overline\nu_e \leftrightarrows \overline\nu_s$ channels.
Two different types of terrestrial experiments place constraints on $\nu_e \leftrightarrows \nu_s$ and 
$\overline\nu_e \leftrightarrows \overline\nu_s$ transitions, reactor 
experiments and short baseline accelerator experiments.  Relevant data of the former comes from CHOOZ 
\cite{Apollonio:1999ae} with it's null result on $\overline\nu_e$ disappearance, and the latter KARMEN, 
\cite{KARMEN 2002}, with a null result on $\overline\nu_\mu \rightarrow \overline\nu_e$ appearance, leading to no
indication of oscillations.  LSND \cite{Athanassopoulos et al.(1998)} is the only short baseline experiment which reported
 a positive signal for $\nu_e \rightarrow \nu_{\mu}$ and this data is used as motivation for a sterile neutrino.

Big Bang Nucleosynthesis(BBN) and the cosmic microwave background (CMB) can provide constraints on the existence of sterile
neutrinos.  The number of new neutrino flavors, which would include a sterile neutrino, is restricted by the actions of the  
relativistic energy density limits during the BBN epoch while simultaneously accounting for the WMAP data, 
e.g. \cite{Cyburt 2005,Barger et al.(2003)}.  This restriction may be lifted through
flexibility in the electron neutrino asymmetry, $\xi_e$, which allows room for the addition of neutrinos beyond the standard
model \cite{Barger (2003)2}.  CMB also places bounds on the neutrino mass; however, these bounds are dependent on the thermal
spectrum of the sterile neutrino \cite{Abazajian et al.(2005)}.  At present, while they provide restrictions, the
 BBN and CMB analysis do not discount light sterile neutrinos.

We organize the paper as follows:  In Section II, we describe the neutrino driven wind model and general nucleosynthesis 
in the 
neutrino-driven wind. Section III describes matter-enhanced active-sterile neutrino oscillations and their effects on
nucleosynthesis.  Section IV describes the network calculation of element
abundances.  In Section V we discuss the 
influence of fission in the $r$process and how fission cycling links the second and third peaks of the $r$process.
Our results appear in Section VI showing our calculation of nucleosynthesis with active-sterile
neutrino oscillations and we compare with both the solar system and halo star abundances.  We conclude in section 
VII.

\begin{figure*}[t]
\begin{center}
$\begin{array}{c@{\hspace{0.5in}}c}
\multicolumn{1}{l}{} &
\multicolumn{1}{l}{} \\ [0.4cm]
\vspace{9pt}
\epsfxsize=2.75in
\epsffile{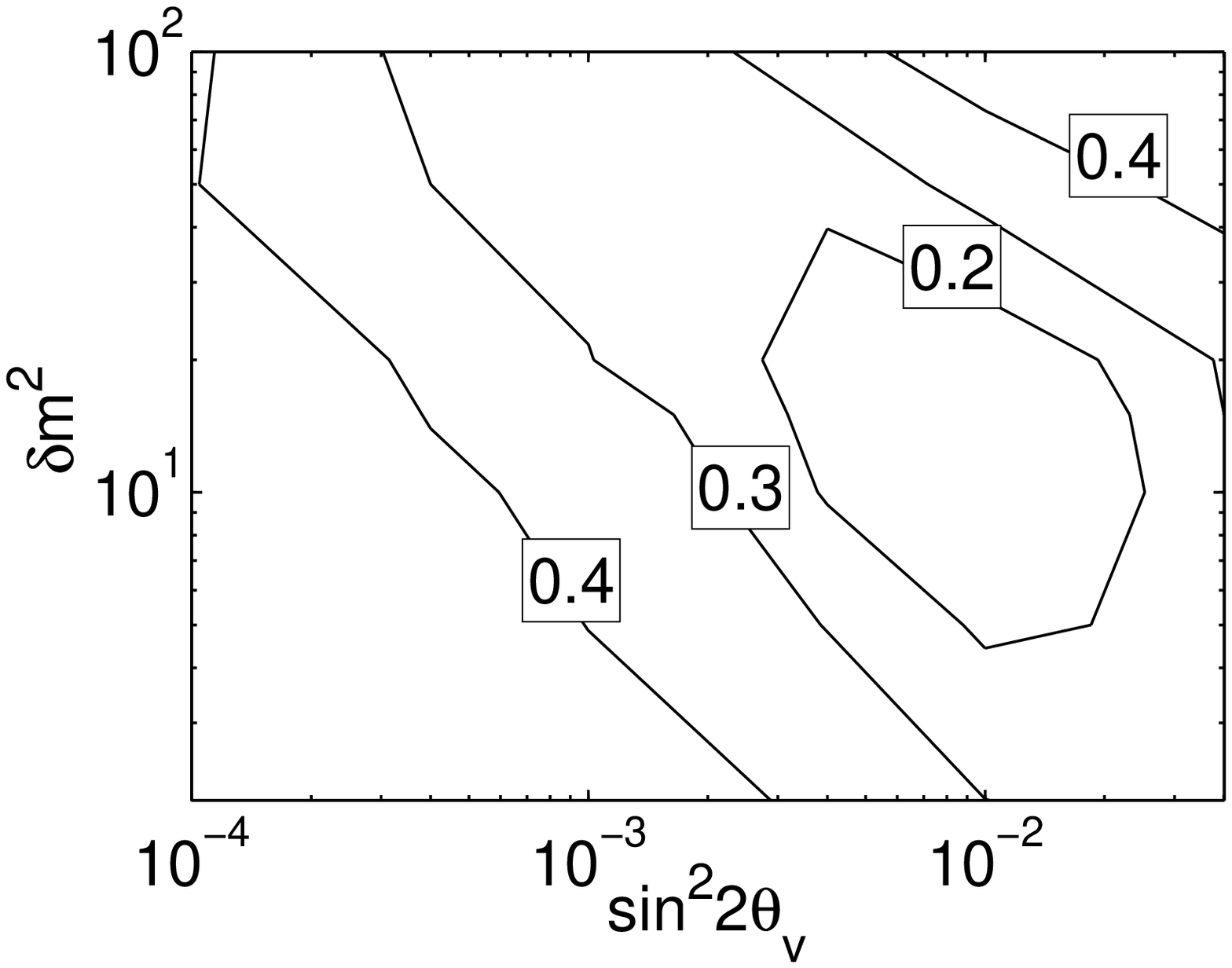} &
\epsfxsize=2.75in
\vspace{9pt}
\epsffile{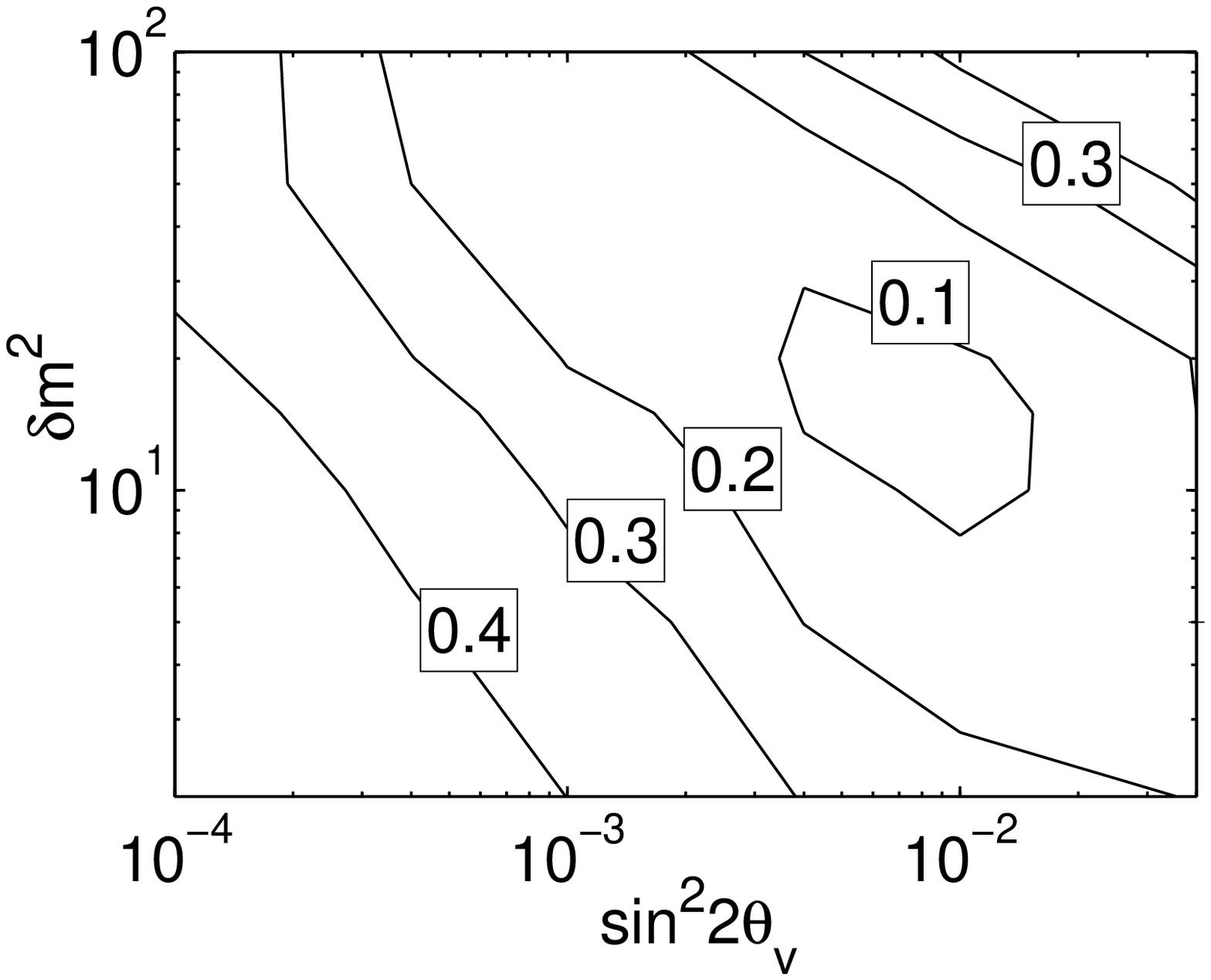} \\ [0.4cm]
\end{array}$
\end{center}
\caption{ In the left panel (Fig. 1a) we show the contours of electron fraction, 
$Y_e$, at the onset of rapid neutron capture, $T_{keV} \approx 200$.
For the astrophysical conditions below, one expects 
an $r$process for $Y_e \lesssim 0.35$ and an $r$process which matches the observed ratio of peaks heights for 
$Y_e \lesssim 0.3$.
For comparison with previous work, in the right panel (Fig. 1b),
 we show the $Y_e$, earlier, at the onset of heavy element formation, $T_{keV} \approx 600 $.
Neutrino captures on free nucleons increase the electron fraction slightly, 
between $T_{keV} \approx 600$ and $T_{keV} \approx 200$.
The neutrino-driven wind parameters are entropy per baryon of $S/k = 100$, expansion time scale of $\tau = 0.3 \, s$, 
neutrino luminosities of $L_{\nu} = 10^{51}\, \rm{ergs}$ and $L_{\bar \nu} = 1.3 \times 10^{51}\, \rm{ergs}$,
 and neutrino temperatures $T_{\nu} = 3.5 \, \rm{MeV}$ and $T_{\bar \nu} = 4.5 \, \rm{MeV}$.}
\label{fig:yeT9T}
\end{figure*}

\section{Nucleosynthesis in the Neutrino Driven Wind}

The neutrino-driven wind epoch begins several seconds after post-core bounce in the type II supernovae environment.
The newborn protoneutron star undergoes Kelvin-Helmholtz cooling, radiating neutrinos that
can lift material off the star's surface.  This material is likely 
ejected by neutrino heating through charged-current reactions.  
It takes about ten neutrino interactions per nucleon to eject material from the 
surface of the star, and the neutrinos set the electron fraction, $Y_e$ \cite{Qian et al.(1993)}.  

In our calculations we use a one-dimensional
model to follow a mass element's peregrination from a nucleon at the surface of the protoneutron star to 
an ejected $r$-process element.
Although multidimensional phenomena such as convection may operate in this environment, this simple model elucidates 
the effect of 
neutrino flavor transformation on nucleosynthesis.

A parameterization of the neutrino-driven wind in the type II supernovae environment following from 
\cite{Duncan et al.(1986),Qian Woosley(1996),Panov Thielemann(2003),Woosley:ux,Mezzacappa Bruenn(1993),Mezzacappa et al.(1998),Mezzacappa:2000jb} 
results in a constant mass outflow rate ($\dot{M}=4 \pi r^2 \rho v$) 
with a homologous outflow with a radial velocity $v$ of

\begin{equation}\label{e:1a}
v \varpropto r,
\end{equation}

where r is the radial distance from the center of the protoneutron star.  From this we find

\begin{equation}\label{e:1b}
r = {r_o} \exp(t/\tau),
\end{equation}

where $r_o$ is the initial radius and the expansion time scale, $\tau$ is defined as

\begin{equation}\label{e:1c}
\tau = r/v.
\end{equation}

Including $\varpropto 1/r$ scaling for the enthalpy per baryon, the density $\rho$ scales as

\begin{equation}\label{e:1d}
\rho \varpropto \rho_o \exp(-3t/\tau).
\end{equation}

In our calculations we primarily use $\tau = 0.3 \, \rm{s}$ and $S/k = 100$ unless otherwise stated.
Our mass element leaves the surface of the protoneutron star in Nuclear Statistical Equilibrium
(NSE) due to the very high temperature.  At entropies of $S/k = 100$ baryonic content exists 
only as protons and neutrons \cite{Qian Woosley(1996)}.
Electron neutrino and anti-neutrino capture on free nucleons set the electron fraction and at the surface produce
 neutron-rich conditions.
As material flows away from the protoneutron star it cools and 
once material reaches temperatures below $T \lesssim 750$ keV the formation of $\alpha$ particles begins.

During this epoch since neutrino interactions have created a neutron-rich environment, 
all the protons and many neutrons bind into $\alpha$ particles. 
Electron neutrinos capture on the remaining neutrons producing protons, which in turn consume more neutrons to form 
additional $\alpha$ particles.  This drives the electron fraction towards $Y_e = 0.5$ decreasing the number of free neutrons
and creating an environment unfavorable to the $r$process, known as the $\alpha$-effect \cite{Meyer:1998sn,Fuller Meyer(1995)}.

As the mass element continues moving away from the protoneutron star, the lower energies allow $\alpha$ particles
and neutrons to combine into seed nuclei for the $r$process.  Farther away from the protoneutron star,
as the material falls out of NSE, it undergoes a series of quasi-equilibrium 
phases \cite{Meyer:1998sn,Meyer et al.(1998),Hix:1999}.  The large Coulomb barriers eventually become insurmountable and 
charged current reactions drop out of equilibrium, while a large abundance of free neutrons remains.  The mass element 
continues to move outward leaving neutron capture (n,$\gamma$) and photo-dissociation ($\gamma$,n) as the only 
reactions left in equilibrium.  

This $(n,\gamma) \rightleftharpoons (\gamma,n)$ equilibrium phase marks the onset of the $r$process,
as the balance between neutron capture and photo-dissociation determines the equilibrium path of
the $r$process.
The $r$process may produce nuclei unstable towards fission at this time, 
whose fission products may effect the $r$-process abundances.
Once there are not enough free neutrons left to maintain $(n,\gamma) \rightleftharpoons (\gamma,n)$ equilibrium, 
neutron capture and 
photo-dissociation reactions freeze-out allowing the $r$-process nuclei to $\beta$-decay back to stability, 
forming the progenitors to the observed $r$-process elements.  

\section{Matter-Enhanced Active-Sterile Neutrino Oscillations}

McLaughlin et al. 1999 \cite{McLaughlin:1999pd} demonstrated that active-sterile neutrino transformation could circumvent
the problematic issues associated with the $r$process in the neutrino driven wind from the protoneutron star.  
Active-active neutrino transformation as well as active-sterile transformation through different channels has also been 
explored in Caldwell et al. 2000 \cite{Caldwell et al.(2000)} and Nunokawa et al. 1997 \cite{Nunokawa:1997ct}.

Here we consider matter enhanced transformation of neutrinos through the $\nu_e \leftrightarrows \nu_s$ and
$\overline\nu_e \leftrightarrows \overline\nu_s$ channels as described by \cite{McLaughlin:1999pd}. 
We define a sterile neutrino as one with interactions significantly weaker than the normal weak interaction.  These 
interactions must be weak enough as to not contribute significantly to the decay rate of the $Z^\circ$ boson 
\cite{Eidelman 2004}.  Our results are independent of the model which produces a light sterile neutrino, as  
any SU(2) standard model singlet could be a light sterile neutrino.

For the following equations we use $\Psi_e$ for the electron neutrino wavefunctions and 
$\Psi_s$ for the sterile neutrino wavefunctions.  We consider evolution of flavor 
eigenstates in matter without the potential created by background neutrinos in the form \cite{Mikheyevetal:1985}

\begin{equation}\label{as5}
i\hbar \frac{\partial}{\partial r}
\left[
\begin{array}{c}
\Psi_e(r) \\
\\
\Psi_s (r)  \end{array} \right]
= \left[ \begin{array}{cc}
\phi_e(r) & \sqrt{\Lambda} \\
\\
\sqrt{\Lambda} & -\phi_e(r) \end{array} \right] \left[
\begin{array}{c}
\Psi_e(r) \\
\\
\Psi_s (r)  \end{array} \right]
\end{equation}

where 

\begin{eqnarray} \label{as6}
\begin{array}{c} \phi_e(r) = \frac{1}{4E}( \pm 2 \sqrt{2} G_F [ N^-_e(r) - N^+_e(r) -– \frac{N_n(r)}{2} ]E  \\
\\
- \delta m^2\cos2\theta_v )
\end{array}
\end{eqnarray}

and 

\begin{equation}\label{as7}
\sqrt{\Lambda} = \frac{\delta m^2}{4E} \sin2\theta_v 
.
\end{equation}

In equation (\ref{as6}), the positive sign designates electron neutrinos and the negative sign electron anti-neutrinos.
Here, $G_F$ is the Fermi constant and $N^-_e(r)$,$N^+_e(r)$, and $N_n(r)$ represent the number density for the electrons, 
positrons, and neutrons in the medium.  The neutrino mixing angle is $\theta_v$, 
and the vacuum mass-squared splitting term is defined as $\delta$m$^2$ $\equiv$ $m^2_2 –- m^2_1$, where $m_1$ and $m_2$ are 
the masses of the mass eigenstates.

The potential is defined to be proportional to the net weak charge,
\begin{equation}\label{as8}
V(r)\equiv 2 \sqrt{2} G_F [ N^-_e(r) - N^+_e(r) - N_n(r)/2],
\end{equation}
meaning neutrinos of energy
\begin{equation}\label{as9}
E_{res} (r) \equiv \pm \frac{\delta m^2 \cos 2 \theta_v}{V(r)}
\end{equation}
undergo a Mikheyev-Smirnov-Wolfenstein (MSW) resonance at a positive ($\nu_e$) or negative ($\overline\nu_e$) potential.
Noting the electron fraction is
\begin{equation}\label{as10}
Y_e (r) = \frac{N^-_e(r) - N^+_e(r)}{N_p(r) - N_n(r)},
\end{equation}
we can immediately see how $Y_e$ influences the sign of the potential,
\begin{equation}\label{as11}
V(r) = \frac{3 G_F \rho(r)}{2 \sqrt{2} m_N} \left(Y_e - \frac{1}{3}\right).
\end{equation}
Electron neutrinos transform for a positive potential and electron anti-neutrinos for a negative potential.
Above, $\rho (r)$ is the matter density a distance $r$ away from the protoneutron star, $m_N$ is nucleon mass, and 
$N_p(r)$ is the proton number density.  Since the local material is electrically neutral we take $N_p = N^-_e - N^+_e$.

Using the above substitutions, we can rewrite equation 
(\ref{as6}) as
\begin{equation}\label{as12}
\phi_e(r) = \pm \frac{3 G_F \rho(r)}{2 \sqrt{2} m_N}\left(Y_e - \frac{1}{3}\right) - \frac{\delta m^2}{4E} \cos2\theta_v
.
\end{equation}
Active-sterile MSW neutrino resonances occur when $\phi_e = 0$, providing a direct
relationship to $Y_e$ from the neutrino mixing parameters, $\delta m^2$ and $\sin^2 2 \theta_{\nu}$
 and the electron neutrino and anti-neutrino energies.

We calculate equation (\ref{as5}) numerically for neutrino survival probabilities passing through matter above the 
protoneutron star.  We assume no flavor transformations occur until the neutrinos leave the surface of the protoneutron
star, hence all neutrinos reside in the active flavor eigenstates, $\nu_e$ and $\overline\nu_e$, with no sterile mixing,
 $\nu_s$ and $\overline\nu_s$, at this time.  
We begin the neutrino transformation evolution a Fermi-Dirac neutrino and anti-neutrino spectrum.
As the neutrinos depart from the protoneutron star, the evolving survival probabilities from equation 
(\ref{as5}) cause the neutrinos to depart rapidly from this initial spectrum.  
We use electron neutrino and anti-neutrino luminosities of $L_{\nu} = 10^{51}$ ergs and $L_{\bar \nu} = 1.3 \times 10^{51}$ ergs 
respectively, electron neutrino and anti-neutrino temperatures of $T_{\nu} = 3.5$ MeV and $T_{\bar \nu} = 4.5$ MeV respectively,
and an effective chemical potential of zero.  We chose these values to be representative of the general qualitative 
behavior of the expected neutrino flavor transformation in this environment.
For these luminosities, electron neutrinos and electron anti-neutrinos strongly determine the initial electron fraction,
 although additional influence on the electron fraction comes from electron and positron capture through the backward
rates of the following reactions:
\begin{equation}\label{as13}
\nu_e + n \leftrightarrows p + e^-,
\end{equation}
and 
\begin{equation}\label{as14}
\bar{\nu_e} + p \leftrightarrows n + e^+.
\end{equation}

\begin{figure}[h]
\centerline{\includegraphics[height=2.5in]{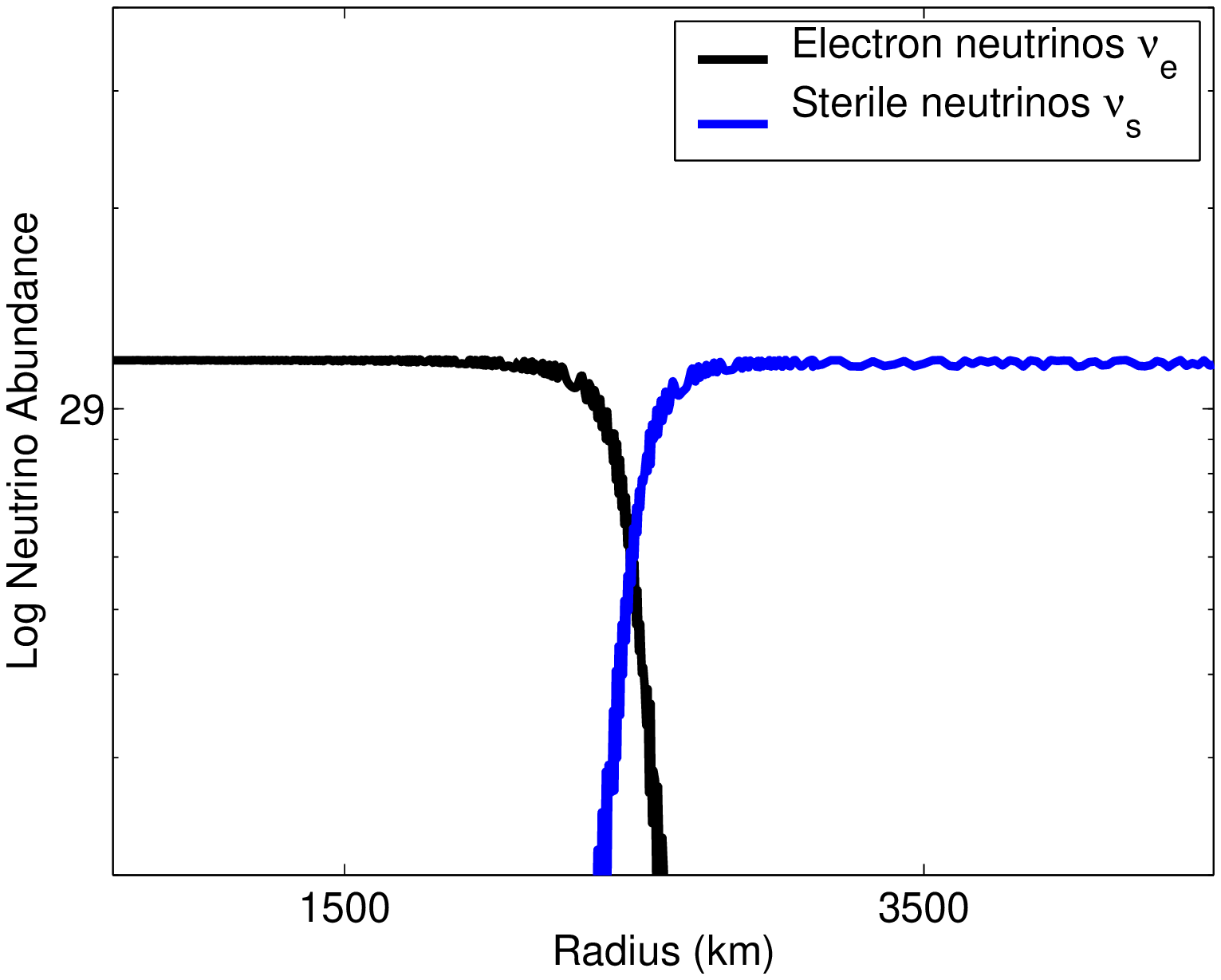}}
\caption{  Electron neutrinos undergoing transformation to sterile neutrinos above the surface of the neutron star
by active-sterile MSW neutrino transformation.
The active-sterile neutrino mixing parameters are $\delta m^2 = 2 \, \rm{eV}^2$ and $\sin^2 2 \theta_v = 7 \times 10^{-2}$,
with the wind parameters of Fig. \ref{fig:yeT9T}. The neutrinos shown have energy $E_{\nu} \approx 11$ keV.  
\label{fig:nu_ab}}
\end{figure}

We now point out the main features of active-sterile neutrino mixing in this environment: the detailed behavior of 
$\nu_e \leftrightarrows \nu_s$ mixing is discussed in \cite{Fetter et al.(2003),McLaughlin:1999pd}.
As neutrinos leave the surface of 
the protoneutron star, low energy electron neutrinos are the first to 
undergo transformation to sterile neutrinos, $\nu_s$, due to the large potential, equation (\ref{as11}).
Such a transformation is shown in Fig. \ref{fig:nu_ab}.
  This reduces 
the efficacy on the forward rate
of equation (\ref{as13}) lowering both the potential and the electron fraction, allowing progressively higher energy
electron neutrinos to transform. Once the electron fraction drops below $1/3$, the highest energy 
electron anti-neutrinos
transform to sterile neutrinos due to the now negative potential.  The potential continues to drop towards its minimum, 
after which it will ascend back towards its zero value.  On its way back, upon the potential reaching this zero value, 
all of the electron anti-neutrinos above a threshold energy convert from $\overline\nu_s$'s  
back to $\overline\nu_e$'s.  
As the mass element reaches the 
alpha particle formation epoch, very few of the original electron neutrinos and most of the original
electron anti-neutrinos are present.  
This severely impedes the forward rate of equation (\ref{as12}) due to the lack of $\nu_e$'s, and also prevents
the dramatic electron fraction rise associated with the $\alpha$ effect, allowing a neutron-rich neutron-to-proton ratio 
needed for
 $r$-process element formation.  
The suppression of the $\alpha$ effect is dependent on the neutrino mixing parameters, 
$\delta m^2$ and $\sin^2 2 \theta_{\nu}$, 
and we describe later which parameters allow for $r$-process elements to be produced \cite{McLaughlin:1999pd}.

\section{Network Calculations}

We calculate the nuclide abundances for a mass element leaving the surface of a protoneutron star 
by linking three network calculations: 
a nuclear statistical equilibrium (NSE) code \cite{McLaughlin:1999pd}, a full reaction rate network code \cite{Hix:1999}, 
and an $r$-process reaction network code \cite{Surman Engel(2001),Surman (1997)}.
  Throughout the calculation we self-consistently solve the neutrino evolution through the integration
of equation (\ref{as5}).  In all regimes we are guided by the structure of the protoneutron star from \cite{Mayle Wilson},
and from the wind as described in \cite{Surman Engel(2001)} as described in Sec. II.  

As the mass element first leaves the protoneutron star with $T_{MeV} \approx 3$, the NSE code takes distance and density 
and calculates all relevant thermodynamic quantities including
positron and electron number densities.  We use NSE to calculate the neutron, proton, and heavy element ($A > 40$) 
abundances.  We also compute the weak reaction rates, generating an updated electron fraction, $Y_e$.  This allows us to 
self-consistently solve the matter enhanced (MSW) equation and hence the neutrino survival probabilities.

Once the mass element leaves the NSE regime, at $T_{keV} \lesssim 900$, we use the full reaction-rate
network to calculate the evolution of the mass element.  In this regime intermediate nuclide production is accounted for,
 $A \lesssim 120$, utilizing strong, electromagnetic, and weak interactions rates from \cite{Rauscher Thielemann(2000)}. 
As before, we continue to use the updated electron fraction to self consistently calculate the MSW neutrino evolution.

As the mass element reaches the neutron capture regime where charged particle reactions have frozen out,
we use an $r$-process reaction rate code, for $T_{keV} \lesssim 200$.
The relevant nuclear reactions included are neutron capture, photo-disintegration, beta decay, charged-current neutrino interactions, and 
beta-delayed neutron emission.  We use the nuclear mass model from \cite{Moller 1995}, 
$\beta$-decay rates from \cite{Moller 1990} and neutron capture rates from \cite{Rauscher Thielemann(2000)}. 
While this mass model is known to produce a gap after the second peak, we choose it since it is the largest self-consistent data set available.
We include the effects of nuclei undergoing fission as it is relevant in this regime.

\section{Fission in $r$-process Nucleosynthesis}

It is important to understand which nuclides become unstable to fission and if and where fission terminates 
the $r$process.  Some previous works such as 
Cowan et al. 1999 \cite{Cowan 1999}
and Meyer 2002 \cite{Meyer(2002)} have included termination of the $r$process through fission.
Meyer 2002 \cite{Meyer(2002)} used a network termination boundary of Z = 91 from proton 
drip line to neutron drip line with only one isotope $^{275}$Pa undergoing fission. 
Cowan et al. 1999 \cite{Cowan 1999} employed a probability of unity for spontaneous fission for nuclei with atomic weight greater than 256.
Other calculations such as Goriely et al. 2004 \cite{Goriely et al.(2004)} and Panov and Thielemann 2004 
\cite{Panov Thielemann(2004)}, employed a set of fission rates over a range of heavy nuclei. 

Perhaps even more important than which nuclei undergo fission is the role the nuclide distribution of the
daughter products play.  There are two fission modes generally considered, 
described by their daughter product distributions, 
symmetric and asymmetric. Symmetric fission results in the progenitor nucleus being split nearly in half by fission,
 resulting in daughter products of the same nuclide for even numbers of protons and neutrons in the progenitor and 
adjacent nuclides for odd permutations of protons and neutrons. For asymmetric fission, the progenitor nucleus splits 
unevenly, resulting in one daughter product being proportionally heavier than the other.

Cowan et al. 1999 \cite{Cowan 1999}
and Rauscher et al. 1994 \cite{Rauscher et al.(1994)} 
employ symmetrical fission, to determine the daughter products, while
Panov and Thielmann 2004 \cite{Panov Thielemann(2004)} utilize a weighting function to determine the various 
daughter products.
Ternary fission has been suggested as well by Ref. \cite{Perelygin 1969,Tsekhanovich 2003}.
At present, a consensus on the choice of fission daughter products has not been formed.

In this work we go one step beyond a single nucleus termination point.  While we terminate the network path at 
$A \geq 270$ by taking the probability of spontaneous fission to be unity above this point, we also include beta-delayed
fission as well for nuclei ranging from $81 \leq Z \leq 100$ and $140 \leq N \leq 164$.
Beta-delayed fission has the greatest importance during late stages of the $r$process \cite{Panov Thielemann(2003)}.
Our approach is sufficient to illustrate the qualitative impact of $\nu_e \leftrightarrows \nu_s$ transform on the $r$process.
For a detailed comparison on a nucleus by nucleus basis with data, it would be preferable to include neutron-induced 
fission as well.  In the future, such rates may become available \cite{Goriely private}.

As stated above, the choice of fission daughter products can have a significant impact on the distribution of 
final $r$-process element abundances.  We examined two models, symmetric fission where the two daughter products are 
taken to be as close to atomic weight and number as possible, 
and asymmetric fission using daughter products suggested by Ref.
\cite{Seeger et al.(1965)}, which takes the daughter products to be about 40\% and 60\% of the mass of the
progenitor nucleus.  

\begin{figure}[h]
\centerline{\includegraphics[height=2.5in]{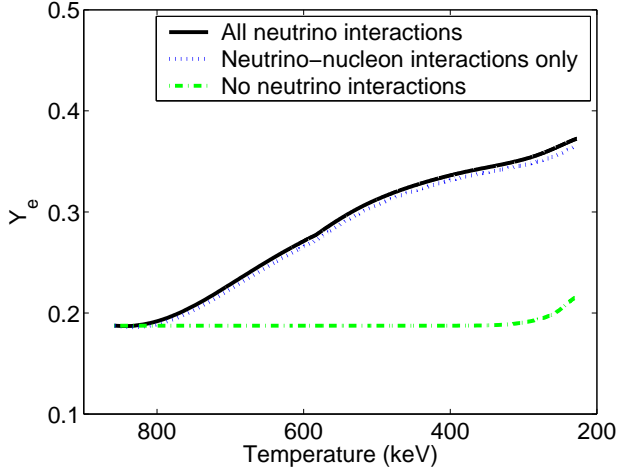}}
\caption{  Electron fraction, $Y_e$ is plotted vs. $T_{keV}$ for the wind parameters in Fig. \ref{fig:yeT9T},
with active-sterile neutrino mixing parameters of $\delta m^2 = 10 \, \rm{eV}^2$ and $\sin^2 2 \theta_v = 0.001$.
The appropriate neutrino interactions are turned off at $T_{keV} = 850$ for the neutrino-nucleon interactions only and
no neutrino interactions cases.
The $Y_e$ continues to increase even after heavy element formation, due primarily to neutrino-nucleon interactions.
\label{fig:yecomp}}
\end{figure}

\begin{figure}[h]
\centerline{\includegraphics[height=2.5in]{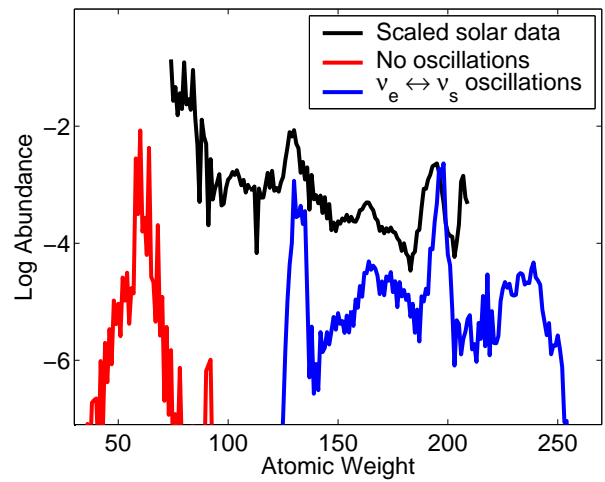}}
\caption{ 
An $r$-process pattern (blue line) producing only the second and third of the three $r$-process peaks
occurs in the neutrino-driven wind when there is active-sterile  
neutrino mixing.
The dramatic impact of active-sterile neutrino oscillations on the $r$process is seen immediately when contrasted against
the abundances produced without neutrino mixing (red line).
For the blue line 
neutrino mixing parameters of $\delta m^2 = 2 \, \rm{eV}^2$ and 
$\sin^2 2 \theta_v = 7 \times 10^{-2}$ are used together with 
same astrophysical conditions as in Fig. \ref{fig:yeT9T}.
The general features of the $r$-process pattern above $A \gtrsim 130$ found in solar system data \cite{Kappeler et al.(1989)} (black line)
are reproduced when $Y_e \lesssim 0.3$ at the onset of neutron capture element formation.
The solar data is scaled to the simulation such that the sum of the $A=195$ nuclides is $2.3 \times 10^{-3}$.
\label{fig:ab}}
\end{figure}

\begin{figure}[h]
\centerline{\includegraphics[height=2.5in]{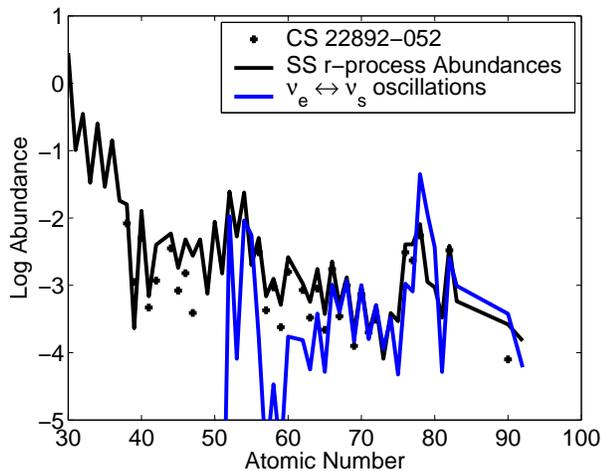}}
\caption{
The general features of the $r$-process pattern found in halo stars are reproduced in the neutrino-driven wind when
active-sterile neutrino oscillations occur.
Only the second, $A \approx 130$ ($Z \approx 55$),  third peaks, $A \approx 195$
($Z \approx 80$), as well as the rare earth bump inbetween, are reproduced.  
The halo star CS 22892-052 \cite{Burris(2000)} abundances are plotted for 
comparative purposes (dots) along with the solar system $r$-process abundances from \cite{Kappeler et al.(1989)}
(black line). 
Neutrino mixing parameters of $\delta m^2 = 2 \, \rm{eV}^2$ and $\sin^2 2 \theta_v = 7 \times 10^{-2}$ were used
for the neutrino driven wind abundances (blue line), although the pattern shown is fairly insensitive to the exact
choice of mixing parameter (see Fig. \ref{fig:peaks}).
The astrophysical conditions are the same as in Fig \ref{fig:yeT9T}.
The measured abundances are scaled to the simulation such that the sum of the $Z = 70$ isotopes in each abundance curve
have an abundance of $10^{-3}$.
\label{fig:halo}}
\end{figure}

\begin{figure*}[t]
\begin{center}
$\begin{array}{c@{\hspace{0.5in}}c}
\multicolumn{1}{l}{} &
\multicolumn{1}{l}{} \\ [0.4cm]
\vspace{9pt}
\epsfxsize=2.75in
\epsffile{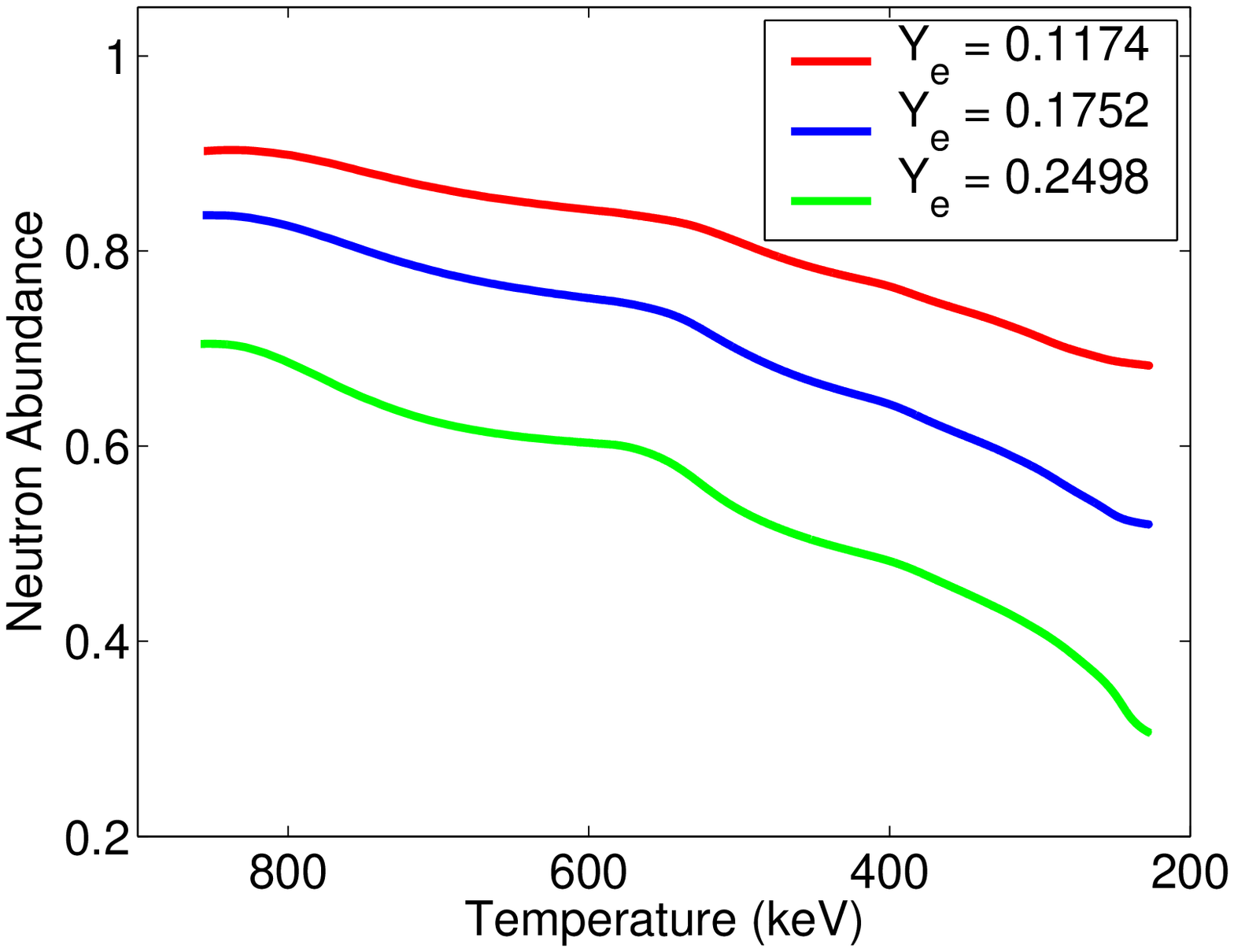} &
\epsfxsize=2.75in
\vspace{9pt}
\epsffile{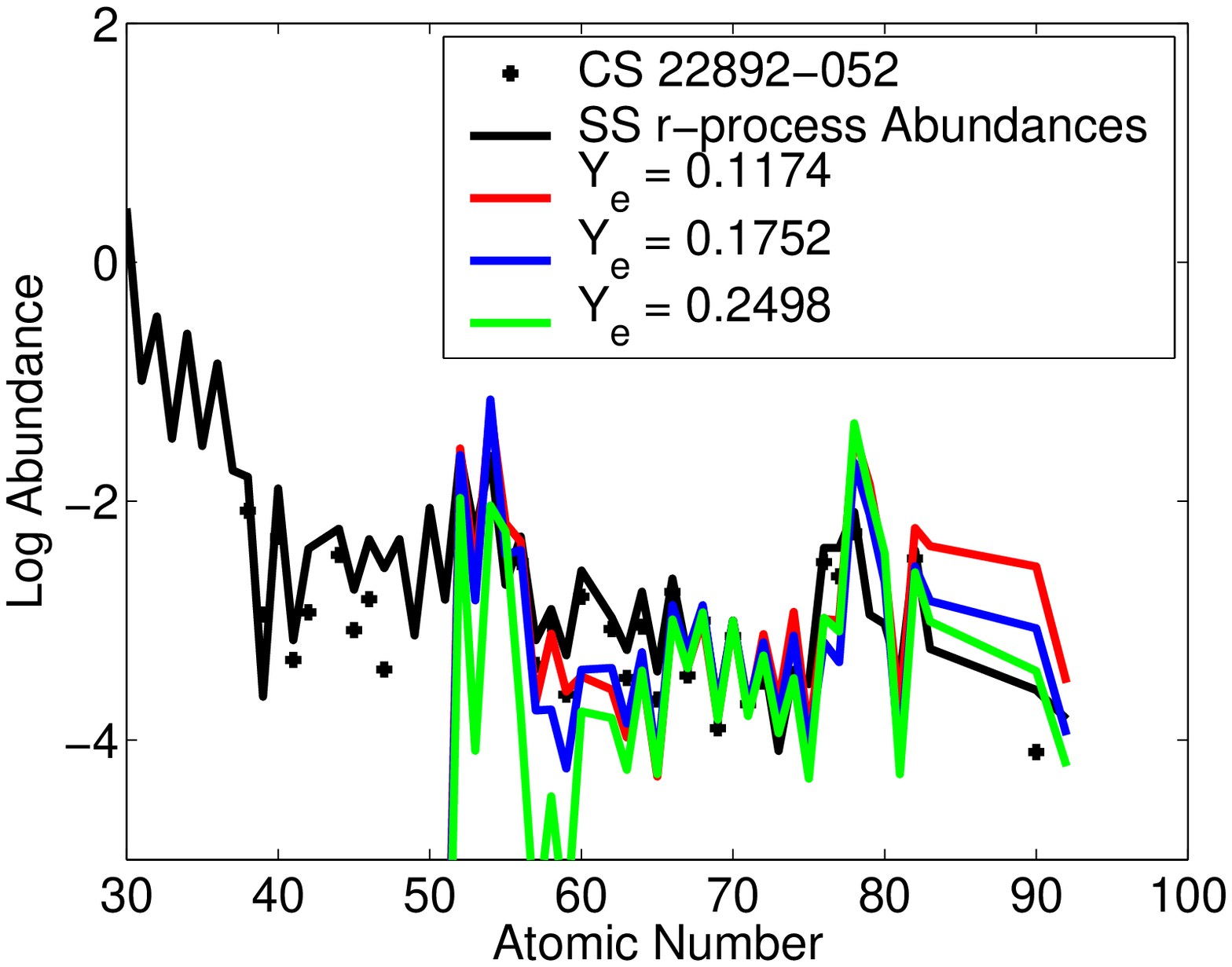} \\ [0.4cm]
\end{array}$
\end{center}
\caption{ Fission cycling produces a consistent overall $r$-process pattern in the neutrino-driven wind,
as long as $Y_e \lesssim 0.3$ at the onset of neutron capture element formation (right panel).
In the left panel we show neutron abundance vs. $T_{keV}$ for three sets of neutrino mixing parameters,  
$\delta m^2 = 10 \, \rm{eV}^2$ and $\sin^2 2 \theta_v = 10^{-2}$,
$\delta m^2 = 5 \, \rm{eV}^2$ and $\sin^2 2 \theta_v = 10^{-2}$, and
$\delta m^2 = 2 \, \rm{eV}^2$ and $\sin^2 2 \theta_v = 7 \times 10^{-2}$ 
corresponding to electron fractions of $Y_e = 0.1174$, $Y_e = 0.1752$, and $Y_e = 0.2498$ respectively. 
The $Y_e$ in the legend occurs at the start of heavy element formation, $T_{keV} \approx 200$. 
Lower $Y_e$'s lead to more neutrons 
but a similar abundance pattern. 
In the right panel we show the abundance patterns from the same mixing parameters as used in the left panel.
The abundances curves and astrophysical conditions are the same as figure \ref{fig:halo}.}
\label{fig:peaks}
\end{figure*}

\begin{figure}[h]
\centerline{\includegraphics[height=2.5in]{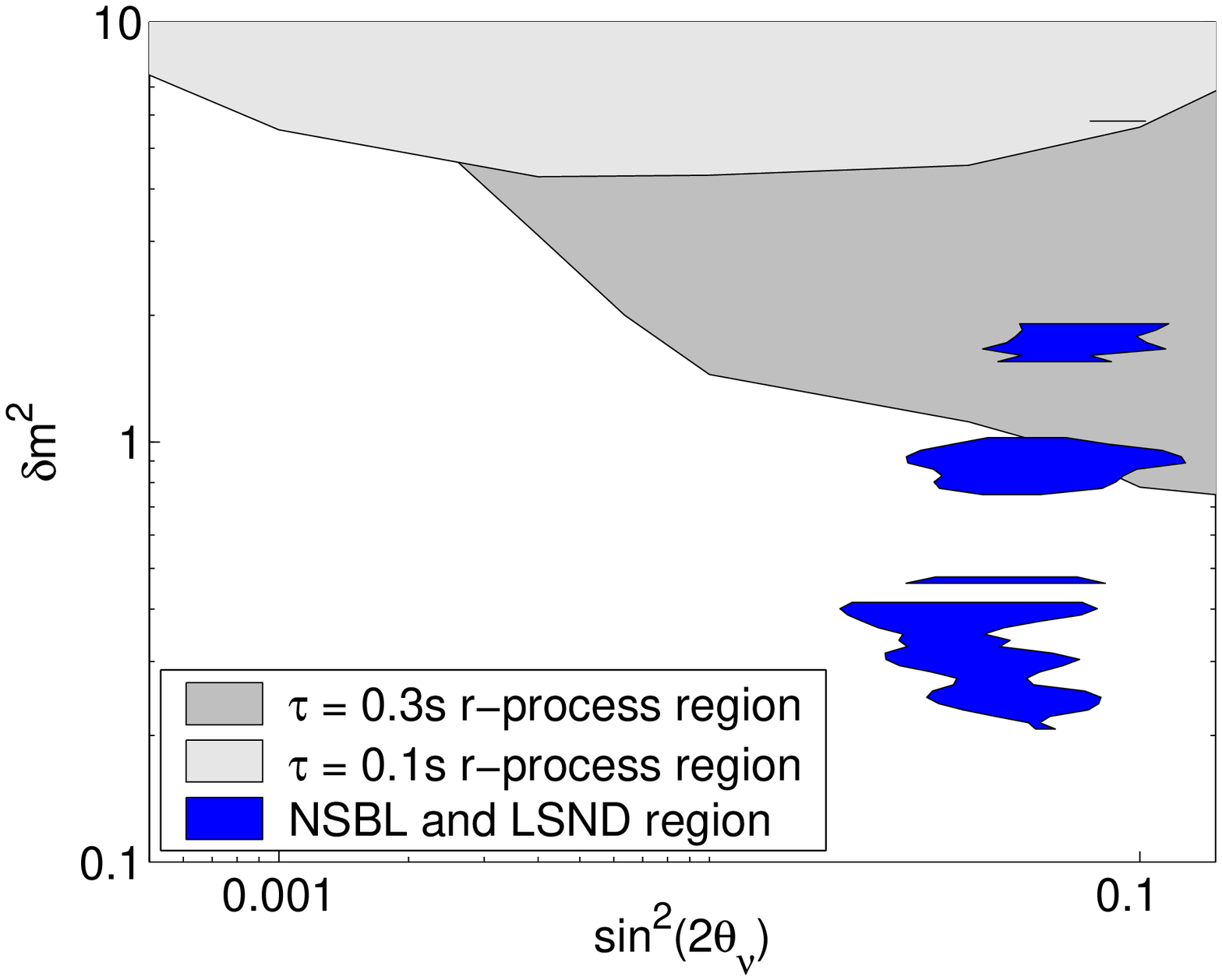}}
\caption{ Contours indicating the parameter space for which we find an $r$process (defined as $Y_e \lesssim 0.35$) 
in the prospective neutrino mixing parameter range of Mini-BooNE.  
The light gray region signifies an $r$process for an outflow timescale of $\tau = 0.1 \, s $, and similarly the dark gray
region for $\tau = 0.3 \, s$, with a reminder of the astrophysical conditions listed in Fig. \ref{fig:yeT9T}.  
The blue region is the combined NSBL and LSND $90\%$ CL allowed region \cite{Ma:2005eh}. 
The parameter space which produces a successful $r$process is much larger than the space probed by experiment.
\label{fig:lsnd}}
\end{figure}


\section{Results and Discussion}

It is convenient to cast our results in the form of both the electron fraction, $Y_e = p / ( n + p)$,
and abundance patterns to allow for direct comparison with
previous theoretical and experimental results.  We begin with a calculation of electron fraction,
$Y_e$, since this is a key 
indicator of $r$-process element production in the neutrino-driven wind environment and it facilitates comparison with 
previous work over the neutrino mixing parameter region, $\delta m ^2$ and $\sin ^2 2 \theta_{\nu}$.
We then discuss the implications of our results and compare our calculated abundance patterns with both solar data
\cite{Kappeler et al.(1989),Burris(2000)} and halo star data \cite{Sneden(2003)}.  We also compare neutrino mixing parameters 
favorable to the $r$process against the parameter region that will be probed by Mini-BooNE \cite{Ma:2005eh}.
We pay particular attention to the effect of fission cycling since this process is a determining factor in the behavior
 of the system.

\subsection{Variation of Neutrino Mixing Parameters}

A key indicator of whether the $r$process will occur is the electron fraction, $Y_e$, 
at the onset of neutron capture element formation.
As described earlier, the $\alpha$ effect together with low entropy 
is a prominent cause of the reduction of free neutrons, stifling the
$r$process.  Active-sterile neutrino oscillations allow the $\alpha$ effect to be circumvented during the epoch of 
$\alpha$ particle formation as electron neutrinos are converted to their sterile counterparts, preventing electron
neutrino capture on neutrons, and  producing a neutron-rich environment.

Our reaction rate network self-consistently accounts for the evolution of $Y_e$ throughout the duration of nucleosynthesis
 of the mass element.  We begin by comparing $Y_e$ with a previous calculation, Fig. 7 of Ref. \cite{McLaughlin:1999pd},
at the onset of heavy element formation (Fig. \ref{fig:yeT9T}b) and find that good agreement is reached.
Then we further track the $Y_e$ up to the point of neutron capture element formation, $T_{keV} \approx 200$, 
and we find that for $Y_e \lesssim 0.35$ at this time, the environment is sufficiently neutron-rich for a 
$r$process to occur (Fig. \ref{fig:yeT9T}a).

We note there is a slight increase of $Y_e$ between the two times, as shown between the two panels of 
Fig. \ref{fig:yeT9T}.
Since charged-current reactions affect the $Y_e$ as the mass element moves farther away from the protoneutron star, 
the increase in $Y_e$ must be the result of one or more charged-current reactions, 
electron neutrino capture on neutrons (the forward rate of equation (\ref{as13})) increases $Y_e$,
making the environment less neutron-rich.  Although electron neutrinos that undergo
transformation to sterile neutrinos do not affect the $Y_e$,  not all electron neutrinos undergo transformation.
The effect of these residual electron neutrinos is shown in Fig. \ref{fig:yecomp}, 
where we plot the electron fraction 
between heavy nucleus production and neutron capture element formation, for the 
neutrino mixing parameters of $\delta$m$^2 = 10 \, \rm{eV}^2$ and $\sin^2 2 \theta_v = 0.001$
Electron neutrino capture on neutrons is therefore responsible for the increase in $Y_e$
seen in Fig. \ref{fig:yeT9T}.

\subsection{$R$-process Abundance Patterns}

In Fig. \ref{fig:ab},
we examine the consequences active-sterile neutrino oscillations have on abundance patterns 
in the neutrino-driven wind environment and
find conditions suitable for the $r$process in this environment 
for $Y_e \lesssim 0.35$. We note that none of our calculations
reproduce the first $r$-process peak at $A \sim 80$.  The resulting abundance pattern features 
the second and third $r$-process peaks and we note that the deficit after the second $r$-process peak, 
$A \sim 130$ is a consequence of the mass model we are using, Fig. \ref{fig:ab} (blue line).
The solar abundance of A = 195 is scaled to $2.3 \times 10^{-3}$ for comparison. 

We classify our abundance patterns by $Y_e$ at the onset of 
neutron capture element formation \cite{Meyer:1997}.
For $0.35 \gtrsim Y_e \gtrsim 0.30$, there are too few free neutrons to produce the third $r$-process peak,
$A \sim 195$, with the same proportions as the second $r$-process peak, $A \sim 130$, as is indicated by observational data.
Electron fractions of $Y_e \lesssim 0.3$ lead to conditions sufficiently neutron-rich to have the second
and third $r$-process peaks in similar proportions to the observational data.
$Y_e$ lower than $Y_e \leq 0.25$ produces $r$-process elements around $A \gtrsim 250$ 
that are strongly susceptible to fission, which can in turn repopulate regions at and around the $A \sim 130$ peak.
We will return to this fission cycling in the following section.
Fig. \ref{fig:halo} shows our abundance pattern plotted as a function of proton number, further demonstrating
the second and third $r$-process peak production.  The three abundances are scaled such that $Z = 70$ has
an abundance of $10^{-3}$.

Early calculations reproduced the general features of the solar system $r$-process pattern by constructing
 a final $r$-process abundance pattern through a mass weighted average of individual abundances patterns 
resulting multiple mass trajectories \cite{Woosley:ux,Meyer et al.(1992),Takahashi et al.(1994)}.
We have investigated different wind conditions, including changes in neutrino luminosity and outflow timescale.
The effect of varying the outflow timescale is shifting the range of 
neutrino mixing parameters, $\delta m^2$ and $\sin^2 2 \theta$,
for which a successful $r$process occurs.
The effect of reducing the neutrino luminosity is contracting the range of 
the neutrino mixing parameters yielding a successful $r$process.
While we find the absence of an $A \sim 80$ peak for a wide range of conditions, it would be necessary to 
calculate a time integrated history over a particular series of wind conditions to determine in a given neutrino
driven wind model produces a significant first $r$-process peak.

\subsection{Effects of fission cycling}

The choice of neutrino mixing parameters affects the free neutron abundance and hence the electron fraction 
at the time of neutron capture element formation.  We find that fission cycling is a mechanism for
generating a stable $r$-process pattern once a threshold $Y_e$ is reached, $Y_e \leq 0.25$ in our environment.
All neutrino mixing parameters which are below this $Y_e$ produce the same general $r$-process pattern,
 shown in Fig. \ref{fig:peaks}.  

Fission cycling behaves as a mechanism linking the second and third peaks of the $r$process.  As very heavy 
nuclides are produced, $A \gtrsim 250$, they may become unstable towards fission and provide an effective termination
point to the $r$-process path.  
The resulting daughter nuclides from this fission repopulate abundances of the second $r$-process peak.

The consistency of the $r$-process pattern is dependent on the equilibrium path of the $r$process.
Early studies analyzing the mass numbers of both the peaks and the closed neutron shells,
yielded a $r$-process path well away from the neutron-drip line.  In general the equilibrium path of the $r$process
is determined by the balance between neutron capture and photo-dissociation for particular temperature and density conditions.
In our calculations the environment becomes sufficiently neutron-rich to generate a $r$-process path along the neutron-drip line, 
resulting in a consistent $r$-process abundance pattern of the second and third peaks.

In the remainder of this section we discuss some details about the effects of fission.
It is difficult to make any definitive comments
on the specific effects that spontaneous fission and beta-delayed fission have on particular 
final $r$-process nuclide abundances due
to uncertainties in fission models; however, some general comments are warranted.
The impact of fission is greatest in the regimes around the $A \sim 130$ peak and past the $A \sim 195$ peak, 
the regimes where fission deposition and depletion occurs.

The details of the pattern around the $A \sim 130$ peak is caused by the deposition of fission daughter products,
which in turn are strongly affected by the choice of fission daughter product distribution,
in the region of $120 \lesssim A \lesssim 140$.
Two modes of daughter product distributions that we examined are symmetric and asymmetric 
as described in Ref. \cite{Seeger et al.(1965)}
and \cite{Chechtkin et al. 1988}.  Symmetric fission modes generally produce an abundance pattern similar to that
produced by an $r$-process calculation without fission, since material is mostly deposited near the closed neutron shell 
region of the $A \sim 130$ peak; however, asymmetric modes can 
make significant material deposits into the $A \sim 100$ region and erase the deficit past the $A \sim 130$ peak. It has
been suggested that fission could form elements $A \leq 120$ in metal poor stars previously, 
\emph{e.g.} Ref. \cite{Panov et al.(2001)}.

Abundance structure of the depletion region past the $A \sim 195$ peak is strongly affected by several factors.
  These factors are the amount of 
material that reaches this region, how strongly fission depletion occurs, and for what elements it occurs.
Most previous work on fission in the $r$process involves calculations which 
effectively terminate the $r$process using a $100\%$ fission probablity in the regime of 
$260 \lesssim A \lesssim 280$.  Given that the nuclear physics of such neutron-rich nuclei is not well understood,
it is difficult to draw conclusions about any detailed features in this region.    
Nevertheless, fission cycling produces a distinct and stable abundance pattern with 
overall features that are in agreement with data.

\subsection{Comparison with Halo Star Abundances}
Halo stars, by virtue of their ancient nature, give us a glimpse of the abundances from the earliest $r$-process progenitors.
The similarity between the $r$-process pattern in halo stars and that seen in the solar $r$process suggests a single
extremely consistent $r$process
mechanism is operating for the heavier $r$-process elements.  
We present our abundance pattern for comparison with the 
halo star CS 22892-052 \cite{Sneden(2003),Kappeler et al.(1989)}  and the solar system $r$process \cite{Burris(2000)} 
in Fig. \ref{fig:halo} and Fig. \ref{fig:peaks}.
There is rough agreement between our data, Fig. \ref{fig:halo} (black line),
 and CS 22982-052, Fig. \ref{fig:halo} (dots),
between $50 \lesssim Z \lesssim 80$ the region of the second and third $r$-process peaks.
  The gap after the $Z \sim 50$ peak is the same as the gap after $A \sim 135$
in the previous plot, shown in Fig. \ref{fig:halo} and is feature of the mass model we are using.
Our final abundances are allowed to decay over a timescale of 14 Gyr for consistency 
with the accepted age of CS 22982-052, using decay modes 
and lifetimes from Ref. \cite{Audi(2003)}.  Adjusting the decay timescale primarily affects the abundances of 
Th$^{232}$ and U$^{238}$ since 
their half-lives are the the shortest of the long lived quasi-stable elements, nearly the same order as the halo star.  
The remaining quasi-stable elements generally have
decay paths to nearby stable nuclides so their decay does not produce a noticeable effect to the abundance pattern.

\subsection{Comparison with Combined NSBL and LSND Data}

The Mini-BooNE experiment is designed to study neutrino mixing in the range of the positive LSND result.
One possible implication of this result is the existence of a sterile neutrino.  In such a scenario
the range of mixing parameters yielding a successful $r$process in our calculations overlaps with the parameter range of Mini-BooNE.
In Fig. \ref{fig:lsnd}, we compare our $r$-process allowed region for two outflow timescales, 
$\tau = 0.1 \, \rm{s}$ and $\tau = 0.3 \, \rm{s}$, with the combined null-short baseline (NSBL) and liquid scintillator neutron detector
(LSND) (3+1) 99\% confidence level (CL) data, with lower CL regions naturally residing inside the 99\% region. 
A compilation of the NSBL and LSND data is described in detail in Ref. \cite{Ma:2005eh}.   
It should be noted that several factors impact the location of the $r$-process region and hence any compatibility with
the NSBL and LSND data.  The choice of expansion timescale, $\tau$, alters the electron fraction and shifts the points where 
active-sterile MSW flavor transformation occurs.  

Finally, we caution that the inclusion of the neutrino background terms in the weak potential can 
significantly change the $r$-process abundances \cite{Patel Fuller 2000,Fetter et al.(2003),Pantaleone(1992)}.
  However, at 
late times the luminosity decreases, decreasing the effect of background terms \cite{Fetter et al.(2003),{Baha 2005}}.
In our calculations which do not include the background, we find that a reduced electron neutrino 
luminosity of $L_{\nu} = 5 \times 10^{50} \, \rm{ergs}$ and 
a reduced electron anti-neutrino luminosity of $L_{\bar \nu} = 6.5 \times 10^{50} \, \rm{ergs}$
can still yield a robust $r$-process solution.
Future studies of neutrino transformation fully inclusive of background terms done together a hydrodynamical wind calculation will determine
 the extent to which $r$-process nucleosynthesis products will be produced and also escape the star.
\section{Conclusions}

We have examined nucleosynthesis in the neutrino-driven wind in the post-core bounce supernovae environment.
We find that inclusion of active-sterile neutrino transformation yields an $r$-process pattern where only 
the second and third of the $r$-process peaks are produced
for a broad range of neutrino mixing parameters.
Analysis of data from halo stars has suggested that there are at least two separate types of events which make 
$r$-process nuclei \cite{Sneden Cowan(2003)}.
Further-more the analysis suggests one of these mechanisms operates early in the evolution of the universe and makes 
primarily the second and third $r$-process peaks.

Fission cycling is responsible for the establishment of a stable $r$-process pattern.  As long as many neutrons 
exist during the phase of neutron capture, nuclei are created with sufficiently high mass number 
that they are unstable to fission.  Below a threshold ($Y_e \lesssim 0.3$) in our model the resulting abundance pattern 
is not sensitive to the magnitude of the free neutron abundance and therefore the number of times the fission cycle occurs.

We find our allowed $r$-process region has overlap with the parameter space to be probed by the Mini-BooNE experiment.
  The extent of this overlap 
is dependent on uncertainties in the astrophysical environment 
and in the neutrino transformation calculations; however, the existence of a sterile neutrino
in the parameters space that will be probed by Mini-BooNE 
will directly influence element production in core-collapse supernova.
 
\acknowledgements

This work was partially supported by the Department of Energy under contract DE-FG02-02ER41216 (GCM) 
and the Research Corporation under contract CC5994 (RS).  
This work was partially supported by the United States National Science Foundation under contract PHY-0244783 (WRH), 
and through the Scientific Discovery through Advanced Computing Program (WRH).  
Oak Ridge National Laboratory (WRH) is managed by UT-Battelle, LLC, for the U.S. Department of Energy under 
contract DE-AC05-000R22725.  
We acknowledge useful discussions with George Fuller and Jim Kneller.

\end{document}